\documentstyle[preprint,aps]{revtex}
\begin{document}
\title{Scaling and singularities in the entrainment\\ of globally-coupled
oscillators}
\author{John David Crawford}
\address{Department of Physics and Astronomy\\
University of Pittsburgh\\
Pittsburgh, Pennsylvania  15260}
\date{December 6, 1994}
\maketitle
\begin{abstract}
The onset of collective behavior in a population
of globally coupled oscillators with randomly distributed frequencies is
studied for phase dynamical models with arbitrary coupling. The population is
described by a Fokker-Planck equation for the distribution of phases which
includes the diffusive effect of noise in the oscillator frequencies. The
bifurcation from the phase-incoherent state is analyzed using amplitude
equations for the unstable modes with particular attention to the dependence of
the nonlinearly saturated mode $|\alpha_\infty|$ on the linear growth rate
$\gamma$. In general we find $|\alpha_\infty|\sim \sqrt{\gamma(\gamma+l^2D)}$
where $D$ is the diffusion coefficient and $l$ is the mode number of the
unstable mode. The unusual $(\gamma+l^2D)$ factor arises from a singularity in
the cubic term of the amplitude equation.
\end{abstract}


The onset of collective oscillations is a multi-faceted phenomenon of interest
in physics, chemistry, biology and most recently
neuroscience.\cite{stro1,kur,win,win2,gray} One
important class of models describes a collection of $N$ dissipative (limit
cycle) oscillators that have some weak mutual interaction. For sufficiently
weak couplings, the basic form of the uncoupled cycle persists, and the
fundamental effect of the interaction is to alter the frequencies and evolving
phases of the oscillators. When the coupling between two oscillators,
$f(\theta_j-\theta_i)$, is uniform for all oscillator pairs, then the evolution
of the phases is given by
\begin{equation}
\dot{\theta}_i=\omega_i+\frac{K}{N}\sum^{N}_{j=1} f(\theta_j-\theta_i) +
\xi_i(t)\label{eq:gcoupled}
\end{equation}
for $i=1,\ldots,N$. If the oscillators are uncoupled ($K=0$), then the phases
simply evolve according to the unperturbed frequencies $\omega_i$ whose
distribution is described by a density $g(\omega)$ characterizing the
population. This normalized density is taken to have zero mean
($\overline{\omega}=0$); this can always be achieved by changing variables,
${\theta}_i\rightarrow{\theta}_i-\overline{\omega}t$, if necessary. For both
physical and mathematical reasons, it is interesting to include in
(\ref{eq:gcoupled}) the effect of extrinsic white noise $\xi_i(t)$ perturbing
the deterministic phase dynamics; this perturbation is defined by the ensemble
averages $<\xi_i(t)>=0$ and $<\xi_i(s)\,\xi_j(t)> = 2D\delta_{ij}\delta(s-t)$.

The form of the coupling function $f(\phi)$ depends on the description of the
underlying limit cycles and their mutual interaction, and will vary from one
setting to another.\cite{kur} Mathematically, since the coupling $f(\phi)$ is
necessarily $2\pi$-periodic, we describe the general form by its Fourier
expansion
\begin{equation}
f(\phi)=\sum_{n=-\infty}^{\infty}\,f_n\,e^{in\phi}.\label{eq:fexp}
\end{equation}
The simplest nontrivial possibility arises when the coupling is dominated by a
single Fourier component, and the theoretical literature is largely focussed on
the case $f(\phi)=\sin\phi$ since this describes a strictly attractive
interaction between oscillators with different phases. Early work by Kuramoto
showed that, in the absence of noise ($\xi_i(t)=0$), there was a critical
coupling strength $K_c$ above which a population $g(\omega)$ would begin to
show frequency locking and partial ordering in the phases $\theta_j$. This
transition was analyzed by an order parameter $R$ defined as the time average
of $R(t)$:
\begin{equation}
R(t)\,e^{i\psi(t)}\equiv\frac{1}{N}\,\sum_{j=1}^{N}\,e^{i\theta_j(t)}.
\label{eq:op}
\end{equation}
For large $N$ if $K<K_c$, then $R\approx 0$, and for $K>K_c$, $R$ scaled like
$R\sim(K-K_c)^{1/2}$. In the limit $N\rightarrow\infty$, this numerical result
was also obtained analytically from a self-consistent calculation of $R$.  For
a population of identical oscillators ($g(\omega)=\delta(\omega)$) in the
presence of noise, Kuramoto analyzed the system of stochastic equations for the
phases via the resulting Fokker-Planck equation which described the phase
distribution. In this theory, the solution with $R=0$ becomes unstable for
$K>K_c$ and a new state with $R\sim(K-K_c)^{1/2}$ emerges. The value for $K_c$
depends on the noise strength $D$ in this case. Subsequently, work has
generalized the Fokker-Planck approach to treat the phase dynamics for
$f(\phi)=\sin\phi$ in populations with nontrivial frequency distributions
$g(\omega)$. These studies also show a bifurcation to phase ordered states with
the same scaling $R\sim(K-K_c)^{1/2}$.

For couplings more general than $f(\phi)=\sin\phi$,
the properties of (\ref{eq:gcoupled}) are not understood and this is an
interesting subject for several reasons. First, the couplings that are {\em
derived} when a reduction to phase dynamics is actually carried out can easily
have a more complicated structure.\cite{kur,hmm} Secondly, recent results by
Daido indicate that as the form of $f(\phi)$ is modified, the nature of the
scaling exponent $R\sim(K-K_c)^{\beta}$ can change from the value
$1/2$.\cite{dainew} Thus different forms of $f(\phi)$ will correspond to
different univerality classes. Daido specifically considers (\ref{eq:gcoupled})
without noise and applies his ``order function'' formalism devised as a
generalization of
Kuramoto's self-consistent calculation of $R$ in the limit
$N\rightarrow\infty$.\cite{dai2}
His treatment assumes the transition is triggered by the $f_{\pm 1}$ components
of the coupling and analyzes the self-consistent equation perturbatively to
leading nonlinear order. The nature of the solution depends on a certain
expression ``$\mbox{\rm Im}\,(f_{-2}\hat{C})$'' where $\hat{C}$ is a
complicated function of the Fourier components of the order function. If this
expression vanishes then the solution scales in the usual manner with
$\beta=1/2$, but if this expression is non-zero then the solution scales with
$\beta=1$. Thus a coupling with $f_1f_2\neq0$ is predicted to produce
transitions with weaker phase ordering near onset than the transitions
associated with the $\sin\phi$ coupling.  A third motivation for analyzing the
transitions in (\ref{eq:gcoupled}) due to couplings of general form is the
unusual character of the bifurcations found in the Fokker-Planck description of
these phase-ordering transitions: In the absence of noise, the unstable modes
correspond to
eigenvalues emerging from a neutral continuum at onset.\cite{sm,smm,jdc0} This
same feature has been noted in instabilities in other systems such as
collisionless plasmas,\cite{jdc1,jdc2} ideal shear
flows\cite{case2,bdl,cs}, solitary waves\cite{pegowein1,pegwein3},
and bubble clouds.\cite{russo} It is known that in some of these systems the
nonlinear interactions between the unstable modes and the continuum can be
singular in the sense that the amplitude equations for the modes become
singular as the eigenvalue approaches the continuum.\cite{jdc1,jdc2,cs} In
these cases the singularities serve to alter the ``expected'' scaling behavior
of the unstable modes. This is known not to occur for (\ref{eq:gcoupled}) when
$f(\phi)=\sin\phi$,\cite{jdc0} but Daido's results suggest this conclusion may
depend crucially on the form of the coupling.

In this Letter, I analyze the bifurcation from the phase incoherent state with
$R=0$, in the limit $N\rightarrow\infty$, within the framework of the
Fokker-Planck equation\cite{sm,smm,jdc0,sak}
\begin{equation}
\frac{\partial\rho}{\partial t}+\frac{\partial(\rho v)}{\partial \theta}=D
\frac{\partial^2\rho}{\partial \theta^2}.\label{eq:eveqn}
\end{equation}
The density $\rho(\theta,\omega,t)$ is defined so that
$N\,g(\omega)\rho(\theta,\omega,t)\,d\theta\,d\omega$
describes the number of oscillators with natural frequencies in
$[\omega,\omega+d\omega]$ and phases in $[\theta,\theta+d\theta]$. Thus
$\rho(\theta,\omega,t)\,d\theta$ denotes the fraction of oscillators with
natural frequency $\omega$ and phase in $[\theta,\theta+d\theta]$ and must
satisfy the normalization $\int^{2\pi}_0\,d\theta\rho(\theta,\omega,t)=1$
when $g(\omega)\neq0$.
In the limit $N\rightarrow\infty$, the deterministic part of the phase velocity
(\ref{eq:gcoupled}) is expressed as an integral over the population
\begin{equation}
v(\theta,\omega,t)=\omega +K \int^{2\pi}_0\,d\theta'
\int^{\infty}_{-\infty}\,d\omega'
f(\theta'-\theta)\rho(\theta',\omega',t)\,g(\omega'),\label{eq:vel}
\end{equation}
and the coupling $f(\phi)$ is described by its Fourier expansion
(\ref{eq:fexp}).

Equations (\ref{eq:eveqn}) and (\ref{eq:vel}) provide a continuum description
of the oscillator population for which issues of stability and bifurcation can
be analyzed in some detail. In this limit, Kuramoto's order parameter
(\ref{eq:op}) is given by
\begin{equation}
R(t)\,e^{i\psi(t)}=\int^{2\pi}_0\,d\theta'
\int^{\infty}_{-\infty}\,d\omega'\,\rho(\theta',\omega',t)\,g(\omega')
\,e^{i\theta'},
\end{equation}
and the incoherent state $(R=0)$ is described by the uniform distribution
$\rho_0={1}/{2\pi}$;
this distribution is an equilibrium for (\ref{eq:eveqn}) since
$v(\theta,\omega,t)=\omega+Kf_0$ at $\rho=\rho_0$. By setting
$\rho(\theta,\omega,t)=\rho_0 +\eta(\theta,\omega,t)$, the dynamics can be
rewritten for $\eta$:
\begin{equation}
\frac{\partial\eta}{\partial t}={\rm \cal L}\eta +{{\rm \cal
N}({\eta})}\label{eq:dyn}
\end{equation}
in terms of the linear operator
\begin{equation}
{\rm \cal L}\eta\equiv D \frac{\partial^2\eta}{\partial \theta^2} -
(\omega+K\,f_0)\frac{\partial\eta}{\partial \theta} +
\frac{K}{2\pi}\int^{2\pi}_0\,d\theta'\int^{\infty}_{-\infty}\,d\omega'
f'(\theta'-\theta)\,\eta(\theta',\omega',t)\,g(\omega')
\label{eq:lop}
\end{equation}
and nonlinear terms
\begin{eqnarray}
{{\rm \cal N}({\eta})}&=&K
\int^{2\pi}_0\,d\theta'\int^{\infty}_{-\infty}\,d\omega'
\,\eta(\theta',\omega',t)\,g(\omega')
\left[\rule{0in}{0.25in}\eta(\theta,\omega,t)\,f'(\theta'-\theta)-
\frac{\partial\eta}{\partial\theta}(\theta,\omega,t)\,f(\theta'-\theta)
\right].
\label{eq:nop}
\end{eqnarray}
In (\ref{eq:lop}) -- (\ref{eq:nop}), $f'(\phi)\equiv df/d\phi$, and note that
the normalization of $\rho$ implies
$\int^{2\pi}_0\,d\theta\,\eta(\theta,\omega,t)=0$.

The linear stability of $\rho_0$, the onset of linear instability, and the
subsequent nonlinear bifurcation have been previously analyzed in detail for
the specific case $f(\phi)=\sin\phi$.\cite{sm,smm,jdc0} For an arbitrary
coupling $f(\phi)$, the generalization of this stability theory is summarized
here as a prerequisite for the bifurcation analysis. The operator ${\rm \cal
L}$
acts independently on each Fourier subspace $\exp(in\theta)$; consequently the
spectrum can be described by analyzing ${\rm \cal L}\Psi=\lambda\Psi$ for
functions $\Psi(\theta,\omega)=\psi(\omega)\exp(in\theta)$ with $n>0$. In
general this spectrum has both eigenvalues and a continuous component: For each
mode number $n=1,2,\ldots$, there is a line of continuous spectrum at
$\mbox{\rm Re}\,\lambda=-n^2D$; in addition ${\rm \cal L}$ has eigenvalues when
the function
\begin{equation}
{\Lambda_{n}\,({z})}\equiv 1+{K\,f_{n}^\ast}
\int^{\infty}_{-\infty}\,d\omega\frac{g(\omega)}{\omega+Kf_0-z-inD}
\label{eq:Lam}
\end{equation}
has roots. More precisely, if for a particular mode number $n=l$, one finds
$\Lambda_{l}\,({z_0})=0$, then $\lambda=-ilz_0$ is an eigenvalue of ${\rm \cal
L}$ and $\Psi(\theta,\omega)=\psi(\omega)\exp(il\theta)$ is the eigenvector for
$\lambda$ with
\begin{equation}
\psi(\omega)=\frac{-K\,f_{l}^\ast}{(\omega+Kf_0-z_0-ilD)}.\label{eq:efcn0}
\end{equation}
The adjoint operator $({\rm \cal L}^\dagger\,A,B)=(A,{\rm \cal L}\,B)$, defined
via the inner product $(A,B)\equiv
\int_{0}^{2\pi}d\theta\int_{-\infty}^{\infty}
d\omega\,A(\theta,\omega)^\ast\,B(\theta,\omega)$, has a corresponding
eigenfunction $\tilde{\Psi}=\tilde{\psi}(\omega)\exp(il\theta)/2\pi$ where
\begin{equation}
\tilde{\psi}(\omega)=\frac{-g(\omega)}
{\Lambda'_{{l}}\,({z_0})^\ast(\omega+Kf_0-z_0^\ast+ilD)}\label{eq:adjef}
\end{equation}
and ${\rm \cal L}^\dagger\tilde{\Psi}=\lambda^\ast\tilde{\Psi}$. The
normalization in (\ref{eq:adjef}) assumes the root $z_0$ under consideration is
simple; this assumption can be relaxed if necessary but is characteristic of
the codimension one bifurcations.\cite{jdc0}
This adjoint eigenfunction satisfies $(\tilde{\Psi},\Psi)=1$ and defines the
projection, $\eta\rightarrow(\tilde{\Psi},\eta)\;\Psi$, from $\eta$ onto the
$\Psi$ component of $\eta$.

The example of a Lorentzian population,
\begin{equation}
g(\omega)=\frac{\epsilon}{\pi}\,\left[\frac{1}{\omega^2+\epsilon^2}\right],
\label{eq:lorentz}
\end{equation}
provides an instructive illustration. For given values of $(l,K,\epsilon)$, the
solutions to $\Lambda_{l}\,({z_0})=0$ are easily located and one finds that
${\rm \cal L}$ has an eigenvalue
\begin{equation}
\lambda=-l\left[ lD+\epsilon+K\,{\rm Im}(f_l)\right] -ilK[{\rm Re}(f_l)+ f_0]
\end{equation}
whenever $K\,{\rm Im}(f_l)<-\epsilon$.
Since $\epsilon>0$ and $K\geq0$, the occurrence of these modes requires a
coupling such that ${\rm Im}(f_l)<0$. For $f(\phi)=\sin\phi$, this is only
possible for $l=1$, but in general the mode number is not constrained. These
modes are linearly stable when $lD+\epsilon+K\,{\rm Im}(f_l)>0$ and become
linearly unstable for $K>K_c$ where $K_c=-(lD+\epsilon)/{\rm Im}(f_l)$. For
$f(\phi)=\sin\phi$ and $l=1$ this reduces to the familiar result
$K_c=2(\epsilon+D)$.\cite{sm}

For $D>0$, the resulting bifurcation for $K>K_c$, can be analyzed by a center
manifold reduction which yields an amplitude equation describing the
time-asymptotic behavior of the unstable mode.  We introduce this amplitude by
writing $\eta$ in terms of  the
critical linear modes $\Psi$  and the remaining degrees of freedom $S$:
$\eta(\theta,\omega,t)=[\alpha(t)\Psi(\theta,\omega)+ {\rm c.c.}] +
S(\theta,\omega,t)$ where $(\tilde{\Psi},S)=0.$ In terms of $(\alpha,A)$ the
evolution equation (\ref{eq:dyn}) becomes
\begin{eqnarray}
\dot{\alpha}&=&\lambda\, \alpha +(\tilde{\Psi},{{\rm \cal N}({\eta})})
\label{eq:alph1}\\
\frac{\partial S}{\partial t}&=&{\rm \cal L} S+{{\rm \cal N}({\eta})}
-\left[(\tilde{\Psi},{{\rm \cal N}({\eta})})\,\Psi  + {\rm c.c.} \right];
\label{eq:S1}
\end{eqnarray}
The center manifold theorem asserts that any new solutions created by the
bifurcation can be found by assuming $S$ is a function of $\alpha$ and
$\alpha^\ast$:
\begin{equation}
S(\theta,\omega,t)=H(\theta,\omega,\alpha(t),\alpha(t)^\ast)=
\sum_{n=-\infty}^{\infty}H_n(\omega,\alpha(t),\alpha(t)^\ast)\,e^{in\theta}.
\label{eq:scm}
\end{equation}
For these solutions $\eta^c$ we have
$\eta^c(\theta,\omega,t)=[\alpha(t)\Psi(\theta,\omega)+ {\rm c.c.}] +
H(\theta,\omega,\alpha(t),\alpha(t)^\ast)$, and their dynamics is described by
the two-dimensional flow
\begin{equation}
\dot{\alpha}=\lambda\, \alpha +(\tilde{\Psi},{{\rm \cal N}({\eta^c})}).
\label{eq:alph1cm}
\end{equation}

The center manifold dynamics (\ref{eq:alph1cm}) and the function $H$ in
(\ref{eq:scm}) are both
constrained by the symmetries of the problem.  The group ${\rm O}(2)$ is
generated by rotations $\beta\cdot(\theta,\omega)=(\theta +\beta,\omega)$
and reflections $\kappa\cdot(\theta,\omega)=-(\theta,\omega)$
which act on functions $\eta(\theta,\omega)$ in the usual way: for
any transformation $\chi\in{\rm O}(2)$,
$(\chi\cdot\eta)(\theta,\omega)\equiv\eta(\chi^{-1}\cdot(\theta,\omega))$. The
operators ${\rm \cal L}$ and ${\rm \cal N}$ commute with rotations for
arbitrary choices of $g(\omega)$ and $f(\phi)$; in addition if
$g(\omega)=g(-\omega)$ {\em and} $f(\phi)=-f(-\phi)$, then ${\rm \cal L}$ and
${\rm \cal N}$ commute with the reflection $\kappa$. In the latter circumstance
the bifurcation problem has ${\rm O}(2)$ symmetry, otherwise the rotational
symmetry alone corresponds to ${{\rm SO}(2)}$.

The Fourier coefficients
of $H$ for $n>0$ are zero unless $n$ is a multiple of $l$, the mode number of
the instability. Rotational symmetry implies that the non-zero coefficients
have the form $H_l=\alpha\,\sigma h_{l}(\omega,\sigma)$ for $n=l$ and
$H_n(\omega,\alpha,\alpha^\ast)=\alpha^m\,h_{n}(\omega,\sigma)\,\delta_{n,ml}$
for $m=2,\ldots$ where $\sigma=|\alpha|^2$ denotes the basic ${\rm SO}(2)$
invariant. The functions $h_{n}(\omega,\sigma)$ are unconstrained by the
rotations, but must satisfy $h_{n}(-\omega,\sigma)^\ast=h_{n}(\omega,\sigma)$
when reflection symmetry holds. Similarly, rotational symmetry implies the
amplitude equation (\ref{eq:alph1cm}) must have the general form
$\dot{\alpha}=p(\sigma)\alpha$ where  $p(\sigma)$ is a real-valued function if
the reflection symmetry holds; otherwise $p(\sigma)$ is generically
complex-valued.

For $K$ near $K_c$, we expand $p(\sigma)=p_0 + p_1\sigma +\ldots$ and
$h_{n}(\omega,\sigma)=h_{n,0}(\omega)+h_{n,1}(\omega)\sigma +\ldots$ and seek
the leading (and presumably dominant) nonlinear terms in the amplitude equation
(\ref{eq:alph1cm})
\begin{equation}
\dot{\alpha}=\alpha[\lambda+p_{1}|\alpha|^2+\cdots].\label{eq:cmnf2}
\end{equation}
The calculation of $p_{1}$ from (\ref{eq:alph1cm}) yields
\begin{equation}
p_1=-2\pi iKl\left[f_l<\tilde{\psi},h_{2l,0}>+
f_{2l}^\ast<\tilde{\psi},\psi^\ast><g,h_{2l,0}>\right]\label{eq:nfcoeff}
\end{equation}
where the brackets denote an integration over $\omega$:
$<A,B>\equiv\int_{-\infty}^{\infty}
d\omega\,A^\ast\,B$, and the function $h_{2l,0}$ is determined
self-consistently to be\cite{remark}
\begin{equation}
h_{2l,0}(\omega)=\frac{-4\pi lKf_{l}^\ast}{2l(\omega+Kf_0-z_0-i2lD)}
\left[\psi(\omega)-\frac{iKf_{2l}^\ast}{2l\Lambda_{{2l}}\,({z_0})}
\int_{-\infty}^{\infty}\,d\omega'
\frac{g(\omega')\psi(\omega')}{\omega'+Kf_0-z_0-i2lD}\right].
\label{eq:cmcoeff}
\end{equation}
For the special case $f(\phi)=\sin\phi$, the instability arises for $l=1$ and
the $f_{2l}$ component is zero. Then the second terms in (\ref{eq:nfcoeff}) and
(\ref{eq:cmcoeff}) are absent and the results of Ref.\cite{jdc0} are recovered.

When $f_{2l}\neq0$, the new terms change the appearance of the bifurcation
significantly due to the factor $<\tilde{\psi},\psi^\ast>$ in
(\ref{eq:nfcoeff}). From (\ref{eq:efcn0}) and (\ref{eq:adjef}) one sees that
the integrand has poles $\omega_\pm=(\Omega/l-Kf_0)\pm i(l^2D+\gamma)/l$ above
and below the contour along the real axis; here $z_0=(\Omega+i\gamma)/l$ is the
root and $\gamma$ the linear growth rate from $\lambda=-ilz_0$.  The
$\gamma\rightarrow0^+$ limit of $<\tilde{\psi},\psi^\ast>$ in $p_1$ produces a
pinching singularity when $D$ is small or zero, and this singularity
contributes a factor $(\gamma+l^2D)$ to the denominator when the integral is
evaluated:
\begin{equation}
<\tilde{\psi},\psi^\ast>=
-\frac{lf_l\,{\rm Im}(f_l)}{(\gamma+l^2D)\,|f_l|^2\,\Lambda'_{{l}}\,({z_0})}.
\label{eq:sing}
\end{equation}
The remaining integrals $<\tilde{\psi},h_{2l,0}>$ and $<g,h_{2l,0}>$ in
(\ref{eq:nfcoeff}) are well behaved as $\gamma\rightarrow0^+$ since all poles
lie in the same half-plane.

The effect of the singularity $(\gamma+l^2D)$ in (\ref{eq:sing}) is clarified
by scaling the amplitude
\begin{equation}
\alpha(t)\equiv
\sqrt{\gamma(\gamma+l^2D)}\,r(\gamma t)e^{-i\xi(t)}\label{eq:rescaled}
\end{equation}
so that the equations for $r(\tau)$ and $\xi(t)$ from (\ref{eq:cmnf2}) are
nonsingular:
\begin{eqnarray}
\frac{dr}{d\tau}&=&r(\tau)\left[1+{\rm Re}\,(p_1)
\,(\gamma+l^2D)\,r^2+\cdots\right]\label{eq:finite1}\\
\frac{d\xi}{dt}&=&\Omega-{\rm Im}\,(p_1) \,\gamma(\gamma+l^2D)\,r^2+\cdots.
\label{eq:finite2}
\end{eqnarray}
Here $\tau\equiv\gamma t$ is the slow time scale determined by the linear
instability, and the coefficients in (\ref{eq:finite1}) and (\ref{eq:finite2})
are now finite as $\gamma\rightarrow0^+$ even when $D=0$. Assuming that as
$t\rightarrow\infty$ the instability saturates with the mode amplitude tending
to a non-zero limit $r(\tau)\rightarrow r_\infty$, then the magnitude of this
mode $|\alpha_\infty|=\sqrt{\gamma(\gamma+l^2D)}\,r_\infty$ determines the
scaling exhibited by the entrained state. For $(\gamma+l^2D)$ sufficiently
small, the amplitude equation (\ref{eq:finite1}) becomes independent of
$\gamma$ and $D$ and the scaling behavior of the entrained state follows the
explicit dependence shown in (\ref{eq:rescaled}). For $D>0$, there is a
crossover from $|\alpha_\infty|\sim\gamma$ for $\gamma>l^2D$ to
$|\alpha_\infty|\sim\sqrt{\gamma}$ for $\gamma<l^2D$. In the noise-free limit,
this crossover does not occur and the $|\alpha_\infty|\sim\gamma$ scaling
persists as the true asymptotic behavior. Since $\gamma\sim(K-K_c)$ near onset,
these results determine the scaling exponent
$|\alpha_\infty|\sim(K-K_c)^\beta$.
In the special case $D=0$ and $l=1$, these conclusions support Daido's
findings: if $f_2=0$ then the singularity is suppressed and the $\beta=1/2$
scaling occurs; if $f_2\neq0$, then the $l=1$ instability leads to an entrained
state characterized by $\beta=1$.

Several important and related questions remain: When $f_{2l}\neq0$ and the
cubic singularity occurs what are the singularities of the higher order terms
in the amplitude equation? Does the amplitude scaling in (\ref{eq:rescaled})
suffice to control the higher order singularities if they occur? Finally, when
$f_{2l}=0$ are there higher order singularities that alter $\beta$ from the
value $1/2$ predicted by the cubic analysis? These issues will be addressed in
a subsequent paper.\cite{jdcdav}


\begin{references}
\bibitem{stro1} S.H. Strogatz, Norbert Wiener's Brain Waves,
in Lecture Notes in Biomathematics, Vol. 100, ed. S. Levin,
(Springer Verlag, New York, 1994).

\bibitem{kur} Y. Kuramoto, {\bf Chemical Oscillations, Waves, and Turbulence},
 Springer-Verlag, New York (1984).

\bibitem{win} A.T. Winfree, Biological rhythms and the behavior of populations
of coupled oscillators, { J. Theor. Biol.} {\bf 16} (1967) 15-42.

\bibitem{win2} A.T. Winfree, {\bf Geometry of Biological Time},
Springer-Verlag, New York (1990).

\bibitem{gray} C.M. Gray, P. Konig, A.K. Engel and W. Singer, Oscillatory
responses in cat visual cortex exhibit inter-columnar synchronization which
reflects global stimulus properties, {Nature} {\bf 338} (1989) 334-337.

\bibitem{hmm} D Hansel, G. Mato, and C. Meunier, Phase dynamics for weakly
coupled Hodgkin-Huxley neurons, { Europhysics. Lett.} {\bf 23} (1993) 367-372.

\bibitem{dainew} H. Daido, Generic scaling at the onset of macroscopic mutual
entrainment in limit-cycle oscillators with uniform all-to-all coupling, {
Phys. Rev. Lett.} {\bf 73} (1994) 760-763.

\bibitem{dai2} H. Daido, Order function and macroscopic mutual entrainment in
uniformly coupled limit-cycle oscillators, { Prog. Theor. Phys.} {\bf 88}
(1992) 1213-1218.

\bibitem{sm} S.H. Strogatz and R. Mirollo, Stability of incoherence in a
population of coupled oscillators, { J. Stat. Phys.} {\bf 63} (1991) 613-635.

\bibitem{smm} S.H. Strogatz, R. Mirollo and P.C. Matthews, Coupled nonlinear
oscillators below the synchronization threshold: relaxation by generalized
Landau damping, { Phys. Rev. Lett.} {\bf 68} (1992) 2730-2733.

\bibitem{jdc0} J.D. Crawford, Amplitude expansions for instabilities in
populations of globally-coupled oscillators, { J. Stat. Phys} {\bf 74} (1994)
1047-1084.

\bibitem{jdc1} J.D. Crawford, Universal trapping scaling on the unstable
manifold of a collisionless electrostatic mode,  { Phys. Rev. Lett.} {\bf 73}
(1994) 656-659.

\bibitem{jdc2} J.D. Crawford, Amplitude equations for electrostatic waves:
universal singular behavior in the limit of weak
instability, { Phys. Plasmas}, to appear (1995).

\bibitem{case2} K. Case, Stability of inviscid plane Couette flow, { Phys. Fl.}
{\bf 3} 143 (1960).

\bibitem{bdl} R.J. Briggs, J.D. Daugherty, and R.H. Levy, Role of Landau
damping in crossed-field electron beams and inviscid shear flow, { Phys. Fl.}
{\bf 13}  421-432 (1970).

\bibitem{cs} S.M. Churilov and I.G. Shukhman, Nonlinear stability of a
stratified shear flow in the regime with an unsteady critical layer, { J. Fluid
Mech.} {\bf 194} 187-216 (1988).

\bibitem{pegowein1} R. Pego and M.I. Weinstein, { Phil. Trans. R. Soc. Lond.
A} {\bf 340} (1992) 47-94; also in {\bf Differential Equations with
Applications to Mathematical Physics}, W.F. Ames, E.M. Harrel, and J.V. Herod,
eds., Academic Press, Orlando, 1993.

\bibitem{pegwein3} R. Pego, P. Smereka, and M.I. Weinstein, Oscillatory
instability of solitary waves in a continuum model of lattice vibrations,
{Nonlinearity}, submitted, (1994).

\bibitem{russo} G. Russo and P. Smereka, Kinetic theory for bubbly flow I:
collisionless case, {  SIAM J. Appl. Math.}, submitted, (1994).

\bibitem{sak} H. Sakaguchi, Cooperative phenomena in coupled oscillator systems
under external fields, { Prog. Theor. Phys.} {\bf 79} (1988) 39-46.

\bibitem{remark} The calculation of $h_{2l,0}(\omega)$ closely follows the
corresponding calculations in Ref.\cite{jdc0}.

\bibitem{jdcdav} J.D. Crawford and K.T.R. Davies, in preparation.

\end{references}
\end{document}